\documentclass[showpacs,twocolumn,secnumarabic,amssymb, nobibnotes, aps, prd]{revtex4-1}

\usepackage{graphicx}
\begin{document}
\title{Traffic dynamics on dynamical networks: The connection between
network lifetime and traffic congestion}%
\author{Xianxia Yang}
\author{Cunlai Pu}
\email[Email: ]{pucunlai@njust.edu.cn}
\author{Meichen Yan}
\author{Rajput Ramiz Sharafat}
\author{Jian Yang}
%\author{The American Physical Society}%
%\email[Corresponding email: ]{pucunlai@njust.edu.cn}
\affiliation{Department of Computer Science and Engineering, Nanjing University of Science and Technology, Nanjing 210094, China}
%\affiliation{1 Research Road, Ridge, NY 11961}
\date{April 13, 2016}%
\begin{abstract}
For many power-limited networks, such as wireless sensor networks and mobile ad hoc networks, maximizing the network lifetime is the first concern in the related designing and maintaining activities. We study the network lifetime from the perspective of network science. In our dynamic network, nodes are assigned a fixed amount of energy initially and consume the energy in the delivery of packets. We divided the network traffic flow into four states: no,  slow, fast, and absolute congestion states. We derive the network lifetime by considering the state of the traffic flow. We find that the network lifetime is generally opposite to traffic congestion in that the more congested traffic, the less network lifetime. We also find the impacts of factors such as packet generation rate, communication radius, node moving speed, etc., on network lifetime and traffic congestion.
\end{abstract}
%\PACS 89.75.Hc, 89.20.-a, 05.10.-a
\maketitle
%\tableofcontents
%\PACS{89.75.Hc, 89.20.-a, 05.10.-a}
\section{Introduction}
Nowadays, human life is increasingly dependent on many information infrastructures such as the Internet, mobile communication networks, sensor networks, ad hoc networks, and so on. Meanwhile, these technological networks face challenging problems including traffic congestion \cite{1Klopper,2sole,3toroczkai}, cascading failures \cite{4zhao,5Buldyrev,6wang,7crucitti}, errors and attacks \cite{8berezin,9pu,10pu,11albert}, virus spreading \cite{12pu,13shen,14pastor,15pastor,16yang,17pastor}, etc., which have been widely discussed in the network science community. Traffic congestion is recognized as a specific state of traffic flow in the network, where nodes cannot manage to deliver their buffered packets and the total network load increases with time, which is generally constant when there is no congestion. The onset of traffic congestion is when the network achieves the maximum network capacity, which is quantified by the critical packet generation rate \cite{18chen,19adi,20Zhao,21arenas}.
If the real packet generation rate is larger than the critical value, the network will endure traffic congestion problem, otherwise,  the network is under free flow state.

Essentially, the network capacity is mostly determined by the network topological structures. It was found that scale-free networks are more susceptible to traffic congestion than homogenous networks \cite{22yan,23guimera}. The reason is that the heterogeneous node degree distribution of scale-free networks leads to the uneven load distribution, which makes large degree nodes carry a large amount of traffic load and thus is easy to trigger the congestion problem.
 However, many real-world complex networks especially communication networks are scale-free networks \cite{24barabasi}. Researchers proposed various strategies \cite{18chen}, which are classified into the ¡°hard¡± and ¡°soft¡± strategies, to avoid traffic congestion and improve the network capacity of scale-free networks. The hard strategies are about optimizing network topological structures by deleting some links or nodes such as the high-degree-first (HDF) strategy \cite{25liu}, the high-betweenness-first (HBF) strategy \cite{26zhang}, and the variance-of-neighbor-degree-reduction (VNDR) strategy \cite{27huang}, or by adding some links between nodes with long distance or nodes around large degree nodes \cite{28huang}. The limitation of the hard strategies is that in real situation it is costly or unpractical to modify the network topological structures. The soft strategies are various routing protocols more applicable to real-world complex networks. The well known shortest path protocol \cite{29awer}  selects the shortest paths as the transmission routes to transmit information as fast as possible. However, the shortest paths share a few large degree nodes, which are very susceptible to traffic congestion.
The other improved routing protocols consider not only  path length, but also node degree \cite{22yan,31zwu,32wang}, node load \cite{33echenique,34zhang,35ling,36wang,37wang}, memory information
\cite{38huang}, next-nearest neighbors \cite{39tadic,40yang}, etc.  Generally, the more information used in path choosing, the better performance the protocol has, and the larger the computational cost is. In addition, those improved protocols often have tunable control parameters to explore the maximum transmission performance.
Most recently, several researchers studied transport processes on multi-layer \cite{41puc,42zhou,43tan,44nian} or multiplex networks \cite{45du,46du,47zhou} with emphasis on optimizing the network capacity and transmission efficiency.

In network science, the lifetime or survival time of a network has not received enough attention, while it is the biggest concern in many power-limited communication networks such as wireless sensor networks, mobile ad hoc networks, etc \cite{48cardei,49dietrich,50akyi,51cali,52male,53abol}. In addition, previously when discussing traffic dynamics, the network topological structures are usually assumed to be static snapshots, in which the nodes and links are fixed. However, many power-limited communication networks have dynamical topological structures, in which nodes are moving and links can only maintain for a certain while. Recently, Yang et al \cite{54yang} proposed an adaptive routing strategy on dynamic networks, but they didn't discuss network lifetime. In this paper, we study traffic dynamics on dynamic networks with limited power supply. We derive the network lifetime based on the level of traffic congestion. We also study how the factors such as packet generation rate, communication radius, node speed, etc.,  affect the network lifetime and traffic congestion.
\section{The network model}
We generate the dynamic networks following Ref. \cite{54yang}. Initially, we set a $L\times L$ square area with periodic boundary conditions. At time $t=0$, we add $N$ moving nodes into the square area. The coordinates of  node $i$ vary with time $t$, which are given as below:
\begin{equation}
 \left\{
 \begin{array}{rcl}
x_i(t+1)  &=&  x_i(t)+v\cos\theta_i(t), \\
y_i(t+1)  &=&  y_i(t)+v\sin\theta_i(t), \\
\theta_i(t+1)  &=&  \theta_i(t)+\phi_i(t).
\end{array}
\right.
\end{equation}
Where $ x_i(t)$ ($y_i(t)$) is the x-coordinate (y-coordinate) of $i$ at time $t$ and $v$ represents the moving speed, which is a constant value for all the nodes.
$\theta_i(t)$ denotes the direction of $i$ at time $t$. $\phi_i(t)$ represents the change  of move direction of $i$ at time $t$, which is a random value uniformly distributed in the interval $[-\pi/3,\pi/3]$ and is generated for each node independently.  $x_i(0)$ ($y_i(0)$) is randomly selected from the interval $[0, L]$. $\theta_i(0)$ is randomly selected from the interval $[-\pi,\pi]$.  Then, the distance between two nodes $i$ and $j$ at time  $t$ is calculated as follows:
\begin{equation}
l_{ij}(t)=\sqrt{[x_i(t)-x_j(t)]^{2}+[y_i(t)-y_j(t)]^{2}}.
\end{equation}
All nodes have the same communication radius $r$. Two nodes are connected by an instantaneous link (communication channel),  if their instantaneous distance is within $r$. Then, the temporal neighbor set of $i$ contains the nodes in its current communication area.

\section{The traffic model}
In our traffic model, all nodes are identical, which can create, buffer, transmit, and receive packets. Specifically, a node generates packets with rate $\rho$, thus each time there are generally $N\rho$ packets inserted into the network. The destination node of a packet is randomly selected among all nodes excluding the source node. All nodes have infinite queues with the first-in-first-out (FIFO) rule for buffering packets. Each time a node can deliver at most $C $ packets. Each node has $E_0$ units of energy at the beginning. The one-hop delivery of a packet will cost $\Delta E$ units of energy.  If the packet's destination node is a neighbor of the node it currently locates, the packet will be directly delivered to the destination node and then be removed immediately. Otherwise, the current node needs to deliver the packet to the appropriate neighbor node based on the given routing strategy. Assuming that at time $t$ the packet is in node $s$, and the destination node $d$ is not in the communication area of $s$. Then, node $s$ will send the packet to its neighbor $i$ with the following probability:
\begin{equation}
P_{si}(t)=(\frac{E_i(t)}{\sum_{j}E_j(t)})^{1-\alpha}/(\frac{l_{id}(t)}{\sum_{j}l_{jd}(t)})^{\alpha}, \alpha\in[0,1],
\end{equation}
where $E_i(t)$ is the residual energy of $i$ at time $t$. The sums in the equation run over the temporal neighbors of $s$. $\alpha$ is a tunable parameter ranging from 0 to 1. When $\alpha =0$, the probability of selecting neighbor node $i$ is proportional to $i$'s residual energy.  When $\alpha =1$, the probability of selecting neighbor $i$ is inversely proportional to the distance between $i$ and $d$. When $0<\alpha<1$, the node distance and residual energy together determine the next hop node.
 If there are no neighbor nodes currently, $s$ will keep its packets and deliver them later. As time goes by, the residual energy of nodes decreases. Following Ref. \cite{55chen}, we assume that the lifetime of the network is from the beginning until the first-dying node appears.
\section{Network congestion}
Ideally, when the delivery capacity of nodes is infinite, there is no traffic congestion phenomenon. However, the delivery capacity is always limited in real situation, and the traffic congestion occurs when the network cannot manage to deliver the continuously injected traffic.
Previously, traffic flow was considered only with two states \cite{18chen,19adi,20Zhao}. When the packet generation rate $\rho$ is no greater than the critical value, the number of packets $S(t)$ in the network is generally constant after a short transition time, and in this case,  the traffic is under the so-called free flow state. When the packet generation rate $\rho$ is larger than the critical value, the number of packets $S(t)$ will increase all through, which is taken as the traffic congestion state.

 In our dynamic network, we obtain four different traffic states, which are shown in Fig. 1. When $\rho$ is very small, $S(t)$ increases abruptly at the first few time steps, and then generally keeps constant until the network dies, which is the no congestion state,  shown in Fig. 1(a). According to Fig. 2(a), in the no congestion state, generally all nodes can deliver their buffered packets at each time step, thus the number of congested nodes $n_c$ is 0,  but the temporal congestion of nodes is allowed, which is why there are small fluctuations in the results of $S(t)$ and $n_c$. As $\rho$ increases and  surpasses the first critical value $\rho_s$, $S(t)$ increases faster and faster after the transition period until the network dies, which is the ¡°slow congestion¡± state, shown in Fig. 1(b). We can infer from Fig. 2(b) that in the slow congestion state, a fraction of nodes become congested first, and then more and more nodes become congested, but when the network dies there are still some nodes,   which are not congested yet.
 Note that there are ``tilting tails" in the curves for the no and slow congestion states as demonstrated in Fig. 1(a), Fig. 1(b), Fig.2(a), and Fig. 2(b).
 The cause of the tilt-tail effect is that, in  the last several steps, the residual energy of most of the nodes is very low, and according to our routing strategy, the packets will be delivered to a few specific nodes of relatively high residual energy, which makes the traffic loads of those nodes substantially increase. When $\rho$ further increases and surpasses the second critical value $\rho_f$, $S(t)$ increases exponentially and then linearly after a very short transition time, which is the ¡°fast congestion¡± state, as shown in Fig. 1(c).
 We can infer from Fig. 2(c)  that in the fast congestion state, firstly a fraction of nodes become congested, and then the congestion gradually spreads to almost all nodes, and thereafter the number of packets  increases linearly, which is different from the slow congestion state. When $\rho$ is larger than the third critical value $\rho_a$, $S(t)$ almost increases  linearly from the beginning, which is the ¡°absolute congestion¡± state, as shown in Fig. 1(d). We see from Fig. 2(d) that in the absolute congestion sate, almost all nodes are  congested after the short transition time at the beginning. Note that for the fast and absolute congestion states, there are no tilt-tail effects because near the ending of the network, all nodes are congested and the residual energy of nodes has no significant difference. The three critical values, $\rho_s$, $\rho_f$, and $\rho_a$,  are marked in Fig. 3.
 \begin{figure}
\centering
\includegraphics[width=3.2in,height=2.5in]{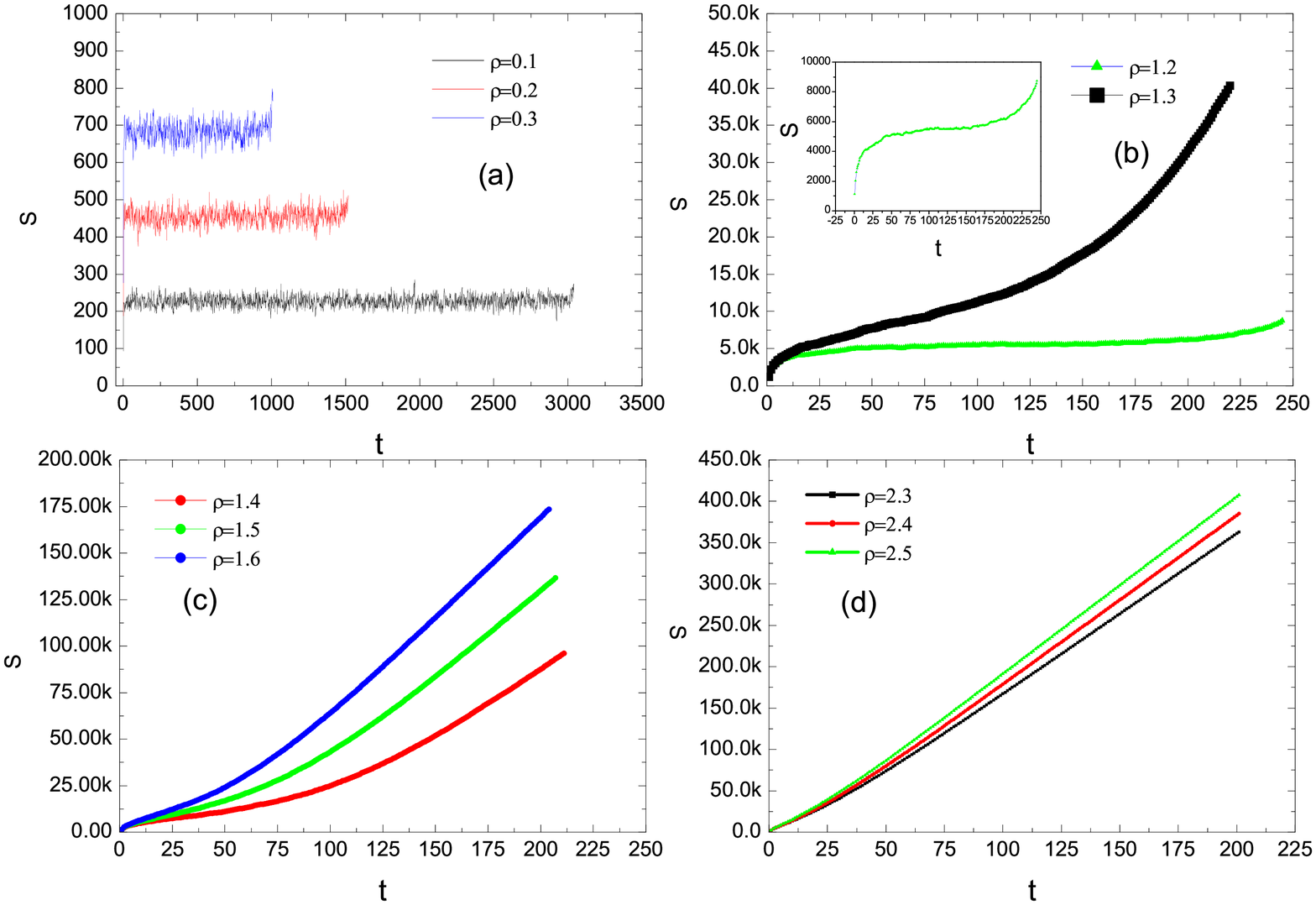}
\caption{$S$ vs. $t$ in different traffic states: (a) no congestion state, (b) slow congestion state, (c) fast congestion state, and (d) absolute congestion state. The simulation parameters are as follows: $N=1000$, $L=20$, $r=3$, $v=0.5$, $\alpha=0.5$,  $\tau_0=3.275$, $C=5$, $E_0=1000$, and $\Delta E=1$. The curves are marked with different colors (see online version). }
\end{figure}
 \begin{figure}
\centering
\includegraphics[width=3.2in,height=2.5in]{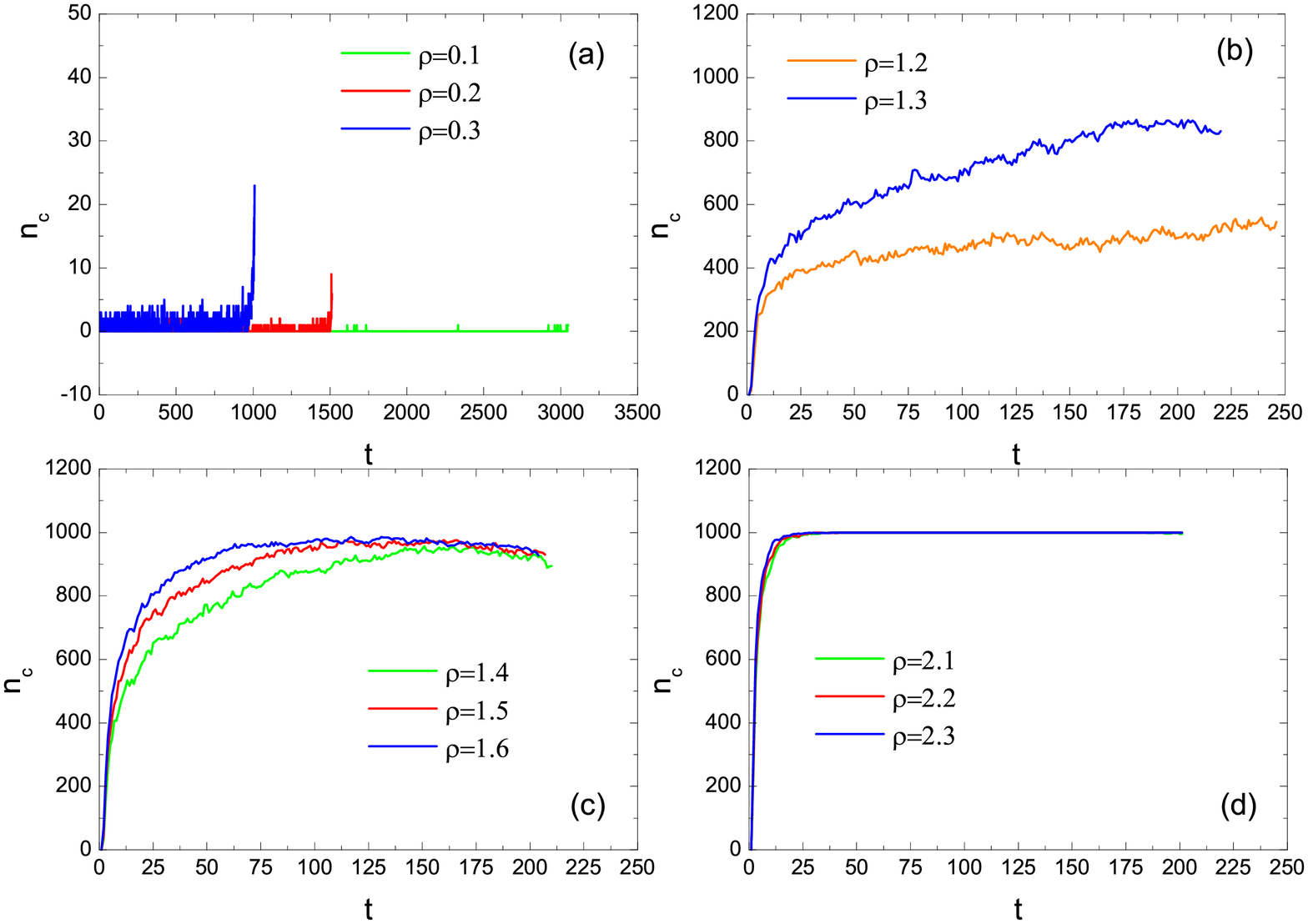}
\caption{$n_c$ vs. $t$ in different traffic states: (a) no congestion state, (b) slow congestion state, (c) fast congestion state, and (d) absolute congestion state. The simulation parameters are as in Fig. 1.  The curves are marked with different colors (see online version).}
\end{figure}
 \begin{figure}
\centering
\includegraphics[width=3in,height=4in]{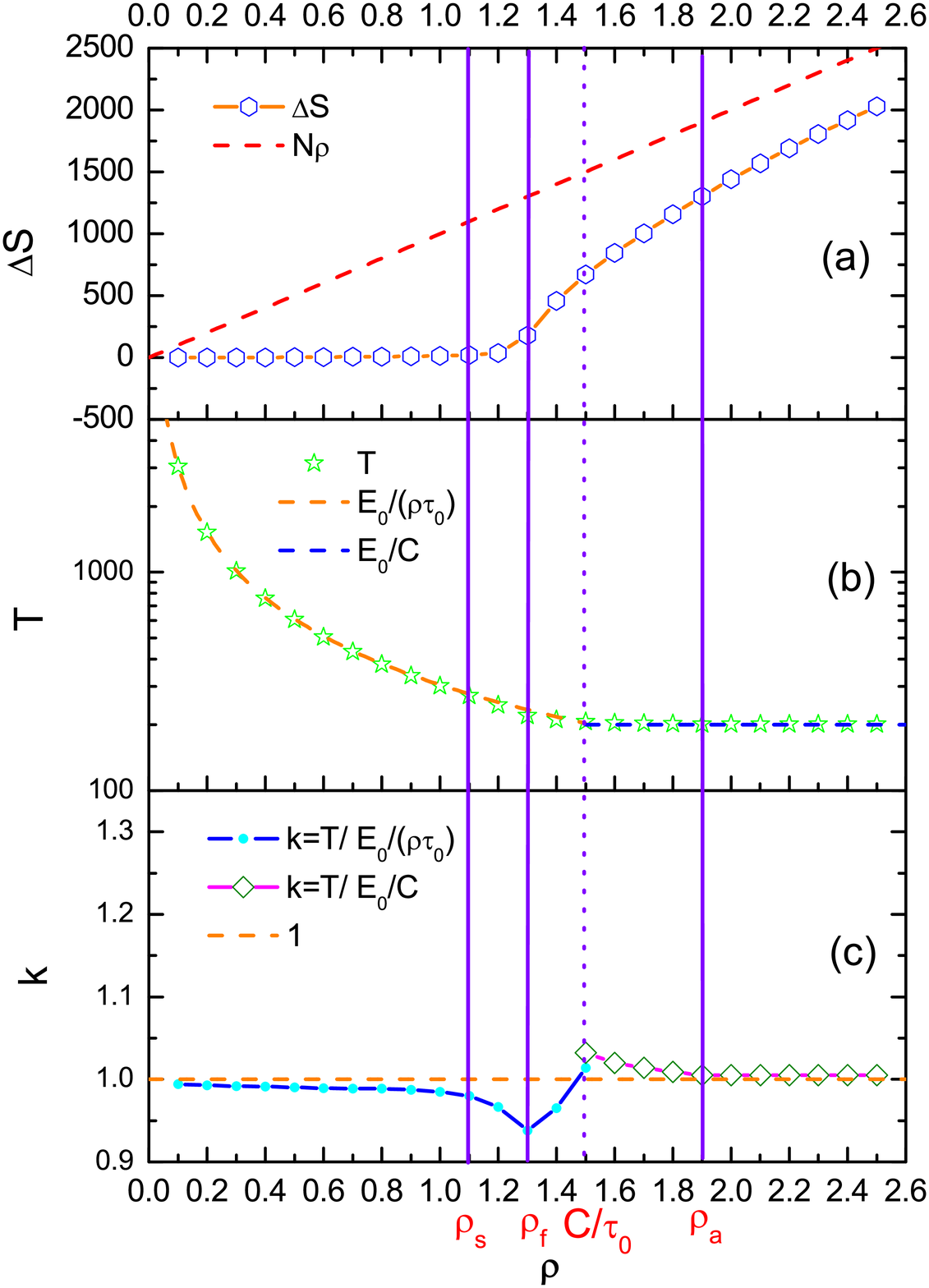}
\caption{(a) $\Delta S$, (b) $T$, and (c) $k$ vs. $\rho$. $\rho_s$, $\rho_f$ and $\rho_a$ are the critical values for slow, fast and absolute congestion states respectively. The simulation parameters are as in Fig. 1.  The results are the average of 1000 independent runs.}
\end{figure}
\section{Network lifetime}
According to Eq. 3, nodes deliver the packets to the neighbor nodes of high residual energy with large probability. In this case, high residual energy nodes consume their energy faster than low residual energy nodes, which leads to a relatively even distribution of the residual energy. We define the range of energy at time $t$ as:
\begin{equation}
R(t)=E_{max}(t)-E_{min}(t).
\end{equation}
Where $E_{max}(t)$ ($E_{min}(t)$) is the maximum (minimum) node residual energy at time $t$.
At the time $T$ when the network dies,
we have $E_{min}(T)=0$ and  $R(T)=E_{max}(T)$. Then, the network lifetime $T$ is calculated as follows:
\begin{equation}
T=\frac{E_{total}(0)-E_{total}(T)}{D*\Delta E},
\end{equation}
where $E_{total}(t)$ is the total units of energy at time $t$. $E_{total}(0)=NE_0$. Since the residual energy is approximately evenly distributed among nodes, $E_{total}(T)\approx N\ast(E_{max}(T)+E_{min}(T))/2=NR(T)/2$. $D$ is the average number of packets delivered each time step for all the nodes.

In the no congestion state, $D=N \rho \tau_0$, where $\tau_0$ is the characteristic transmission time, that is the average number of transmission hops  from source to  destination nodes. Then, the network lifetime $T$ for the no congestion state is calculated as follows:
\begin{eqnarray}
T=\frac{E_{total}(0)-E_{total}(T)}{D\ast\Delta E}
&=&\frac{NE_0-NR(T)/2}{N\rho\tau_0\ast\Delta E}\nonumber \\
&=&\frac{E_0-R(T)/2}{\rho\tau_0\Delta E}.
\end{eqnarray}

In the absolute congestion state, we have $S(t)>D=N\ast C$. Almost all nodes are congested at the  beginning, and every node delivers $C$ packets at each time step, and thus consumes the energy with the same rate. When the network dies,  the total residual energy of  nodes is close to zero, that is $E_{total}(T)\approx0$ or $R(T)\approx0$. Note that when $\rho>C$, all nodes are congested at the very beginning, and in this case, $E_{total}(T)$ and $R(T)$ are definitely 0. Then, the network lifetime $T$ for the absolute congestion state is given below:
\begin{equation}
T=\frac{E_{total}(0)-E_{total}(T)}{D*\Delta E}\approx\frac{NE_0}{NC\Delta E}\approx\frac{E_0}{C\Delta E}.
\end{equation}

For the slow and  fast congestion states, there are periods of exponentially increasing $S(t)$, when $D$ is hard to estimate precisely. In addition, it is not possible to calculate the duration of the tilt-tail effects in the slow congestion state. For all these reasons, we set a nonlinear parameter $k$ to measure the compositive effects of those unpredictable factors on network lifetime $T$. The unified formula of $T$ for all the traffic states is given as follows:
\begin{equation}
T=k\frac{E_0}{\Omega \ast \Delta E}, \Omega =\min\{\rho \tau_0, C\}.
\end{equation}
For the no congestion state, $D=N\rho \tau_0<NC$, that is $\rho \tau_0<C$, we have $k=\frac{E_0-R(T)/2}{E_0}\rightarrow 1$. For the absolute congestion state, $D=NC<N\rho \tau_0$, that is $C<\rho \tau_0$, we have $k=1$. For the slow and fast congestion states, $k$ depends on the nonlinear factors. For the slow congestion state, $\rho \tau_0<C$, then $\Omega=\rho \tau_0$. For the fast congestion state,  $\Omega$ is dependent on $\rho$. When $\rho<C/\tau_0$, $\Omega=\rho \tau_0$, otherwise, $\Omega=C$.

\section{Simulation results}
We study the impacts of factors on traffic congestion and network lifetime through simulation. The key factors of our model include packet generation rate $\rho$, communication radius $r$, node speed $v$, routing parameter $\alpha$, area size $L$, and network size $N$. For traffic congestion, we mainly study the average change rate of number of packets $\Delta S=\overline{S(t)-S(t-1)}$, which measures the average level of traffic congestion during the lifetime of a network.

According to the above analytic results, packet generation rate $\rho$ has significant influence on the traffic congestion and network lifetime. We first study the impacts of $\rho$ by fixing the other parameters as follows: network size $N=1000$,  routing parameter $\alpha=0.5$, node speed $v=0.5$, communication radius $r=3$, area size $L=20$, node delivery capacity $C=5$, and node's initial units of energy $E_0=1000$, per node energy consuming rate $\Delta E=1$. Based on these parameters, we obtain the characteristic transmission time $\tau_0=3.275$.

In Fig. 3(a), we see that in the no congestion state ($\rho\leq\rho_s$), $\Delta S=0$, in the absolute congestion state ($\rho\geq\rho_a$), $\Delta S$   approaches $N\rho$, and in the slow and fast congestion states ($\rho_s<\rho<\rho_a$), $0<\Delta S <N\rho$. In Fig. 3(b), with the increase of $\rho$, $T$ first decreases greatly, and then gradually converges to the minimum value $T=200$. The reason for the results in Fig. 3(b) is that larger packet generation rate means larger number of packets injected into the network, and in this case,  more energy is consumed in each time step, which results in a smaller network lifetime. When the traffic is in the absolute congestion state, the energy consumed in each time step is nearly constant, and thus the network lifetime is a constant value, which can also be inferred from Eq. 7. Moreover, in Fig. 3(b) the analytical and simulation results agree well with each other. In Fig. 3(c), we see that when the traffic is in the no congestion state, $k$ is smaller than but very close to 1. When the traffic is in the absolute congestion state, $k$ is equal to 1. When the traffic is in the slow or fast congestion state, $k$ deviates from 1, which demonstrates the impact of nonlinear factors. These simulation results of $k$ are consistent with the above analytical results.
 \begin{figure}
\centering
\includegraphics[width=2in,height=3in]{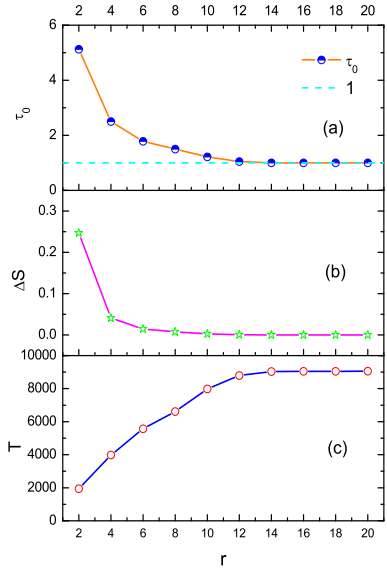}
\caption{(a) $\tau_0$, (b) $\Delta S$, and (c) $T$ vs. $r$. The simulation parameters are as follows: $N=1000$, $L=20$,  $v=0.5$, $\alpha=0.5$,  $\rho=0.1$, $C=5$, $E_0=1000$, and $\Delta E=1$. The results are the average of 1000 independent runs.}
\end{figure}

The other factors, such as communication radius $r$, node speed $v$, routing parameter $\alpha$, etc.,  do not appear in the derivation of network lifetime. However, they affect the characteristic transmission time or the range of energy when there is no traffic congestion, and thus indirectly  affect the network lifetime. Note that when the traffic congestion is heavy, the network lifetime is almost constant and irrelevant of these factors, which can be inferred from Eq. (7).
In Fig. 4, $\Delta S$ is very small, which means that there is almost no traffic congestion. When $r$ is small, $\tau_0$ is relatively large, and thus $T$ is relatively small, which can also be inferred from Eq. (6). With the increase of $r$, $\tau_0$ decreases, and  $T$ increases accordingly. When $r$ is large enough, $\tau_0=1$. In this case, every packet is delivered from the source node to the destination node only by one hop, and $T$ reaches the maximum value.
 \begin{figure}
\centering
\includegraphics[width=3in,height=2.5in]{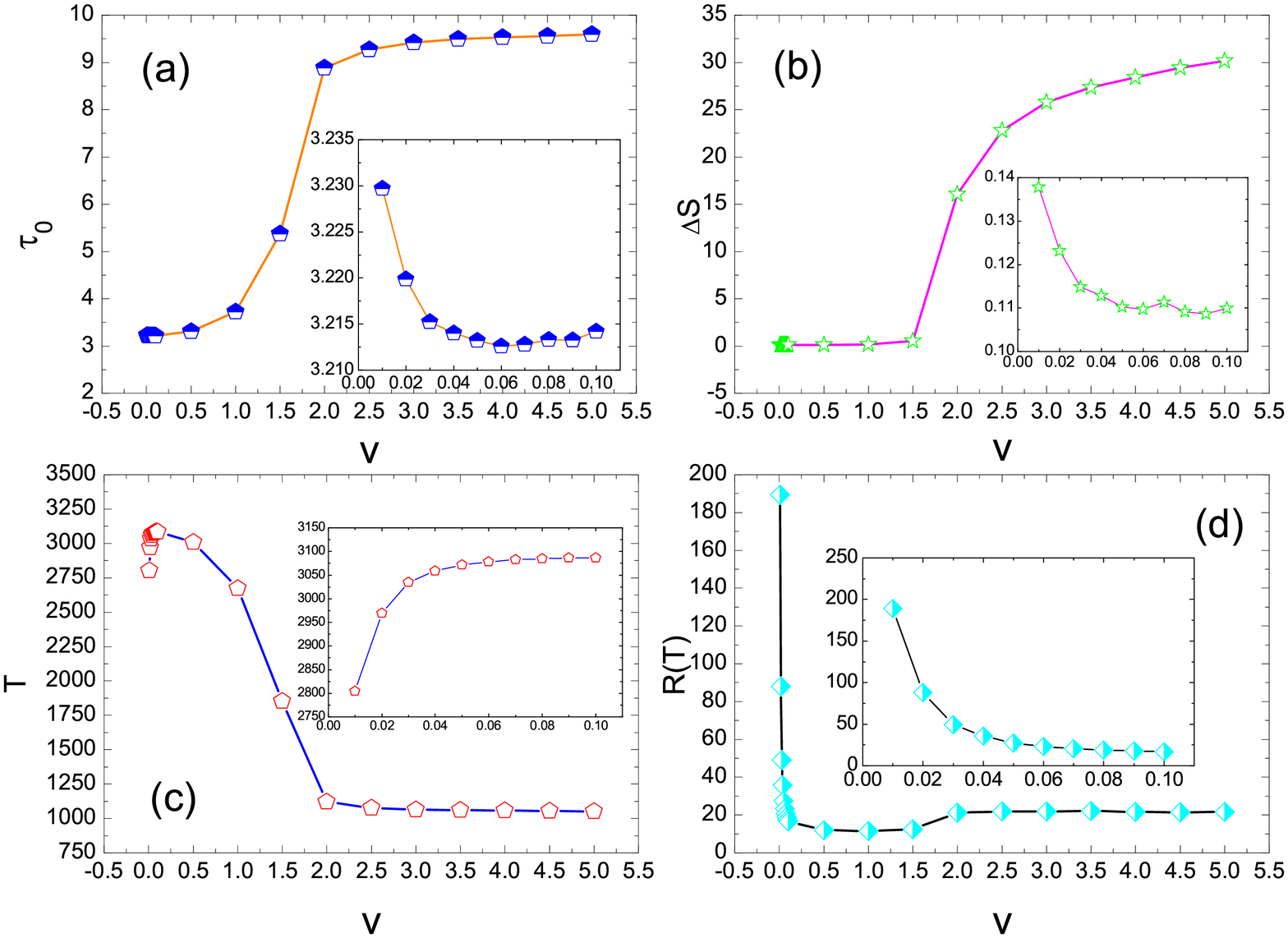}
\caption{(a) $\tau_0$, (b) $\Delta S$,  (c) $T$, and (d) $R(T)$ vs. $v$. The simulation parameters are as follows: $N=1000$, $L=20$, $r=3$,   $\alpha=0.5$,   $\rho=0.1$, $C=5$, $E_0=1000$, and $\Delta E=1$. The results are the average of 1000 independent runs.}
\end{figure}
 \begin{figure}
\centering
\includegraphics[width=3in,height=2.5in]{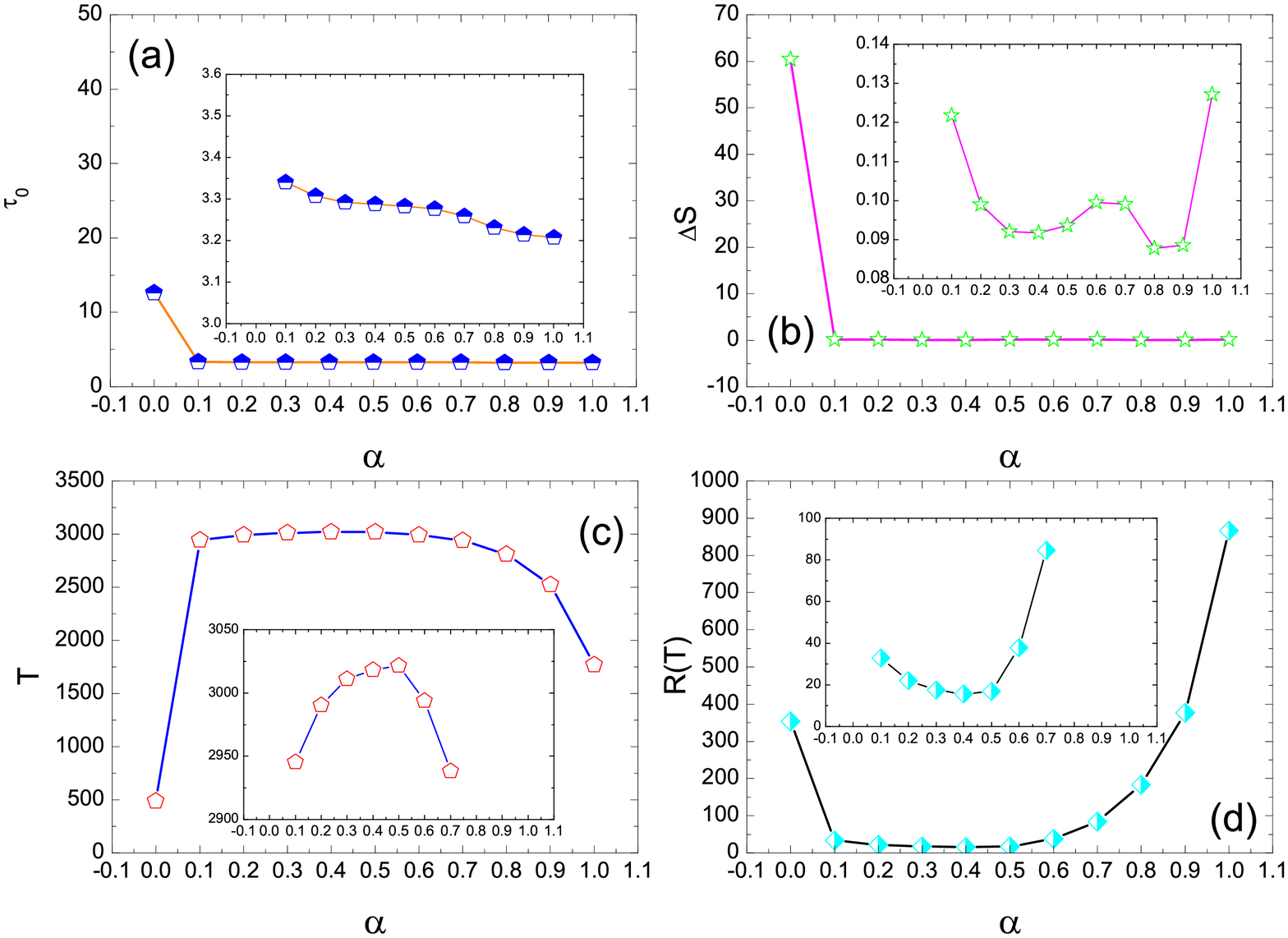}
\caption{(a) $\tau_0$, (b) $\Delta S$, (c) $T$, and (d) $R(T)$ vs. $\alpha$. The simulation parameters are as follows: $N=1000$, $L=20$, $r=3$,  $v=0.01$,   $\rho=0.1$, $C=5$, $E_0=1000$, and $\Delta E=1$. The results are the average of 1000 independent runs.}
\end{figure}
In Fig. 5, when $v$ is very small, $\Delta S$ equals to 0, which indicates that there is no traffic congestion. In this case, $\tau_0$ slightly decreases with $v$, and $R(T)$ also decreases with $v$, both of which lead to the increase of  $T$. As $v$ increases, $\tau_0$ and $\Delta S$ increase accordingly, which cause the decrease of $T$.
In Fig. 6, when $\alpha$ increases from zero to nonzero, $\tau_0$, $\Delta S$ and $R(T)$ decrease abruptly, leading to a substantial increase of $T$.  As $\alpha$ further increases, $\Delta S$ is  around 0,  and $\tau_0$ is almost constant. $R(T)$ slightly decreases with $\alpha$ and then increases, which causes that  $T$ slightly increases with $\alpha$, and then decreases.

Finally, we study how area size $L$ and network size $N$ affect network lifetime. In Fig. 7, when $\Delta S$ is small (the traffic is slightly congested),  $\tau_0$ increases with $L$, and thus $T$ decreases with $L$.  The effect of increasing $L$ is equivalent to decreasing $r$, which can also be inferred by comparing Fig. 7 with Fig. 4.   In Fig. 8, when $\Delta S$ is close to 0 (there is almost no traffic congestion), $\tau_0$ decreases with $N$, and thus $T$ increases with $N$. More nodes results in more paths for packets delivery, which is why the characteristic transmission time becomes smaller with an increasing number of nodes.
 \begin{figure}
\centering
\includegraphics[width=2in,height=3in]{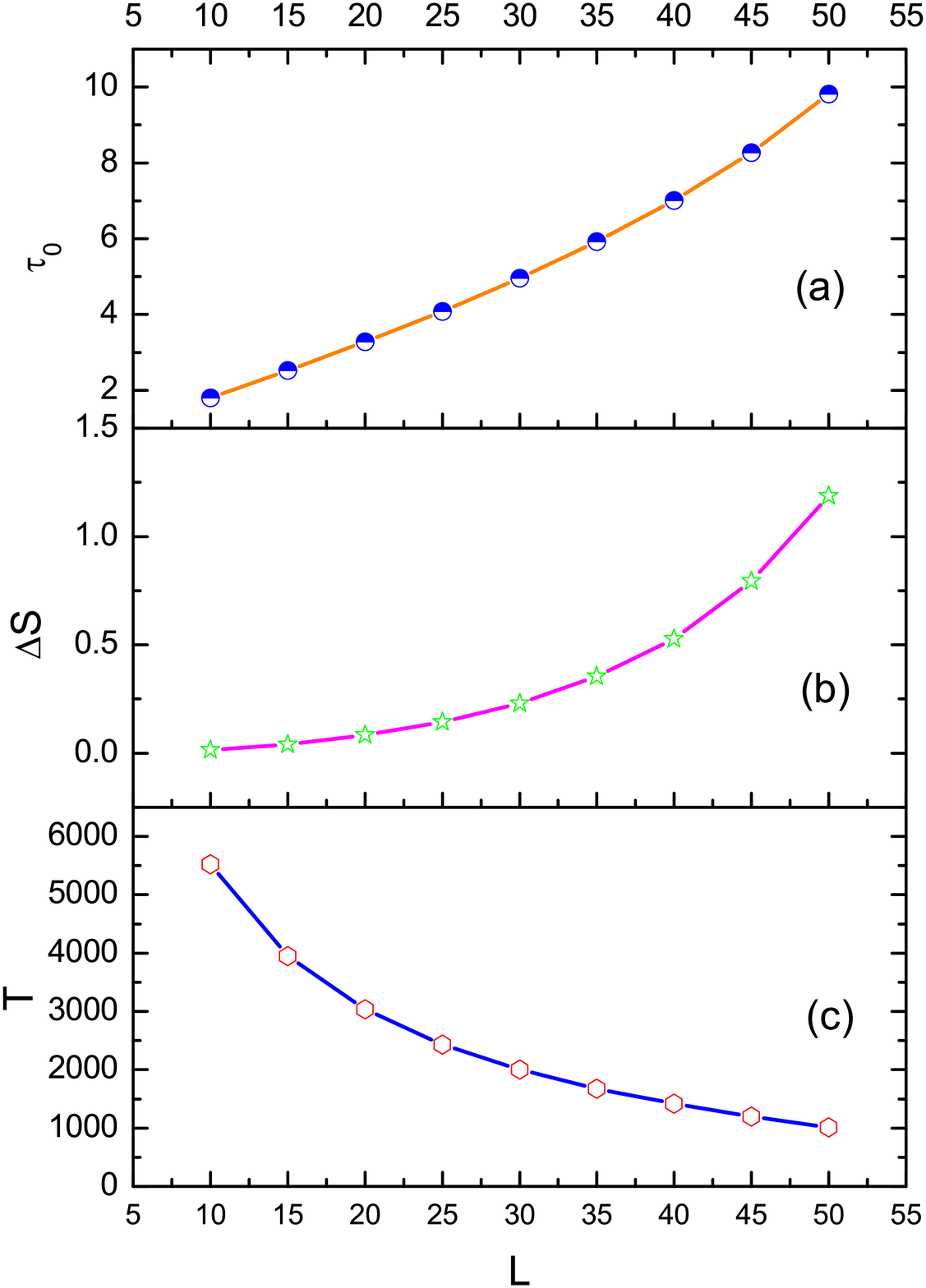}
\caption{(a) $\tau_0$, (b) $\Delta S$, and (c) $T$ vs. $L$. The simulation parameters are as follows: $N=1000$, $r=3$,  $v=0.5$, $\alpha=0.5$,  $\rho=0.1$, $C=5$, $E_0=1000$, and $\Delta E=1$. The results are the average of 1000 independent runs.}
\end{figure}
 \begin{figure}
\centering
\includegraphics[width=2in,height=3in]{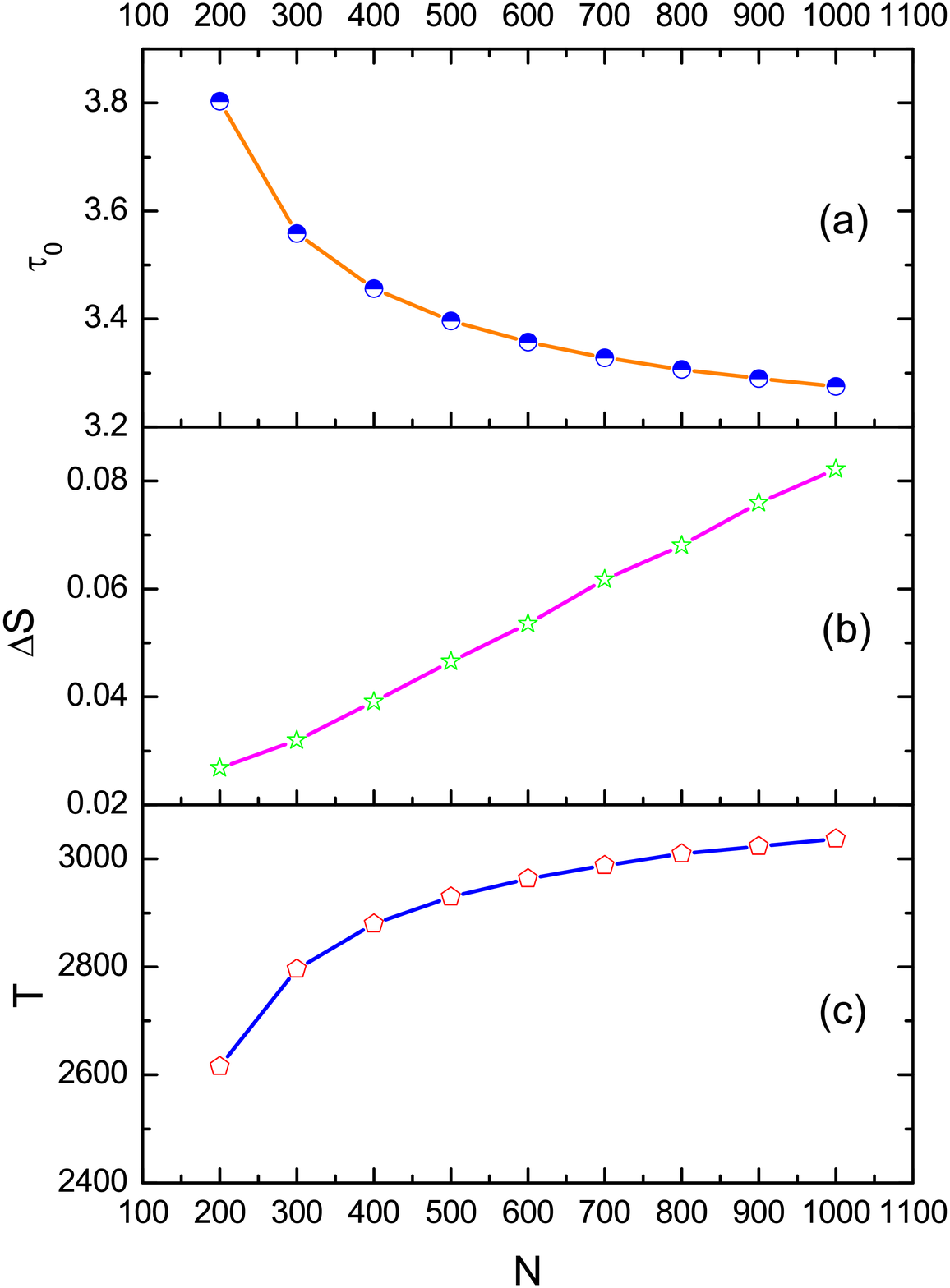}
\caption{(a) $\tau_0$, (b) $\Delta S$, and (c) $T$ vs. $N$. The simulation parameters are as follows:  $L=20$, $r=3$, $v=0.5$, $\alpha=0.5$,  $\rho=0.1$, $C=5$, $E_0=1000$, and $\Delta E=1$. The results are the average of 1000 independent runs.}
\end{figure}

\section{Conclusion}
For a wide range of power-limited communication networks, the most concern is network lifetime, which has not received enough attention in network science. In this paper, we discuss both network lifetime and traffic congestion based on the methodology of complex network theory. In our model, all nodes move in a spatial area and have limited communication radius, energy and delivery capacity, but  the  infinite queue for simplification purpose.
Previously, we only considered if there is traffic congestion in a network. In this paper, we further study the level of traffic congestion, which is divided into no, slow, fast and absolute congestion states. Moreover, we derive network lifetime by considering the level of traffic congestion. Generally, network lifetime is opposite to traffic congestion in that high level of network congestion corresponds to small network lifetime. Through analytic and simulation results, we find that when the traffic congestion is slight, network lifetime is mainly determined by packet generation rate, characteristic transmission time,  and range of energy. When the traffic congestion is heavy, network lifetime is constant and determined by the node delivery capacity.  Also, increasing communication radius decreases the possibility of traffic congestion, and thus increases network lifetime. The influence of routing parameter and node speed is not monotonic in that there are optimal routing parameter and optimal node speed leading to maximum network lifetime.
\section*{Acknowledgments}
This work was  supported by the National Natural Science Foundation of China (Grant Nos. 61201173 and 61304154), the Specialized Research Fund for the Doctoral Program of Higher Education of China  (Grant No. 20133219120032),  the Postdoctoral Science Foundation of China (Grant No. 2013M541673), and China Postdoctoral Science Special Foundation (Grant No. 2015T80556).

\end{document}